\documentclass[12pt]{elsart}
\usepackage{graphicx,amssymb}
\journal{Chaos, Solitons \& Fractals}
\begin{document}
\begin{frontmatter}

\title{Computational universes}

\author{Karl Svozil}
\ead{svozil@tuwien.ac.at}
\address{Institut f\"ur Theoretische Physik,
 University of Technology Vienna,
 Wiedner Hauptstra\ss e 8-10/136,
 A-1040 Vienna, Austria}

\begin{abstract}
Suspicions that the world might be some sort of a machine or algorithm existing ``in the mind''
of some symbolic number cruncher have lingered from antiquity. Although popular at times,
the most radical forms of this idea never reached mainstream.
Modern developments in physics and computer science have lent support to the thesis,
but empirical evidence is needed before it can begin to replace our contemporary world view.
\end{abstract}

\begin{keyword}
Cellular automata \sep digital physics
\PACS 05.65.+b \sep 02.50.-r \sep 02.60.-x
\end{keyword}
\end{frontmatter}

\section{Historical notes}

In a broad context,
the development of rationalism, the enlightenment
and science can be perceived as an
awakening from the illusory world of the senses ({\it Maya} in Sanskrit);
as a growing awareness that "facts" which once were perceived as self-evident
turned out to be utterly wrong.
Humanity once took it for granted that it was located at the epicentre of the Universe.
A closer inspection revealed that there is no ground to claims of any preference in location:
Earth is conveniently situated in a solar system of a remote part of our galaxy,
which in turn is part of a group of galaxies and of the physical
Universe as we perceive it today.
People also trusted that their bodies are made-up of solid stuff.
Later on they  learned that, as their bodies consist of atomic and subatomic ``point'' particles,
things only appear to be solidly filled, but in another perspective, space is ``almost empty.''
Time turned out to be relative to the motion of observers, and
single ``particles'' such as
photons and neutrons, seemed to be at two or more spatial positions at once.
On another issue, people previously thought that they have been created in a
different way than other species. As it turned out, from a biological point of view,
mankind evolved and spread just like locusts and everyone else around.
This is corroborated not only by phylogenetic evidence,
but by analysis of the very DNA code  that constitutes the genetic heritage
and blueprint of our ancestors and of all living beings.
Indeed, the DNA itself turns out to be a biochemical code running on
cellular computers to the effect of creating,
maintaining and reproducing the organism of which it is a part.

Further disillusionments may lie ahead.
Consciousness is still an  ``undiscover'd country,''
and maybe it is just a manifestation of neuronal brain functions.
Or, consciousness may be just the opposite: transcendental.
Despite the achievements of Freud, certain dream phases are barely understood.
Artists have
speculated that we are "fleshware"
units inside of a simulation-computation-game
that appears gigantic or even infinite to us.
Who knows, we might have even paid for to live
a life in the twenty-first century in a beyond fair.
That is to say, we might be embedded in a literal ``game''
that we chose to pass the time.
 To make things more realistic,
all memories of the our life in the
beyond might have been erased from our immediate memories
\footnote{
This is mind-body dualism in a new form.
For a concrete mind-brain interface model, see
for instance Eccles' proposal \cite{eccles:papal}.
In this view, what appears to us as the physical world
is just a simulation-computation-game; and the mind(s) of the player(s) is (are) transcendental
with respect to the characters in this emulation.
Note the phrase {\it ``we are the dead on vacation''} in  Godard's  film
{\it Breathless.}}.
Maybe the ``meaning'' of our world  is rather trivial; like the simulation of marketing
measures for a beyond world
\footnote{
This is the theme in
Galouye's  1964 novel {\it Simulacron 3};
the novel stimulated Fassbinder's {\it Welt am Draht}, as well the recent movie
{\it Thirteenth Floor}.
Somewhat related scripts are those of {\it Total Recall} and
{\it Matrix}.
In {\it Contact,} Sagan  mentions the "Zoo hypothesis" claiming
that there is somebody (in this case aliens) watching us
for ethnographic or other reasons.
Philosophical speculations include
Rene Descartes' {\it world-as-a-lucid-dream} vision \cite[Meditation 1,9]{descartes-meditation},
and Putnam's {\it brain-in-a-vat} metaphor \cite{putnam:81}; see also
{\tt http://whatisthematrix.warnerbros.com}.
}.
As computers have begun to permeate our societies,
it is no wonder that
the ``universe as a computer'' metaphor for the physical
Universe has attracted increasing attention.
Perhaps some day our own technology could achieve such visions,
and put it to our practical use
\footnote{
There is a possibly apocryphal story \cite[p. 127]{Thomas} that, when asked by his Prime Minister
Peel or  by the Chancellor of the Exchequer  Gladstone about the usefulness of his findings,
Faraday responded,
{\it ``Why, sir, there is the probability that you will soon be able to tax it.''}
}.

In antiquity,
 Pythagoras (6th cent. B.C.)
{\it ``considered numbers as the essence and principle of all things,
and attributed to them a real and distinct existence;
so that, in his view, they were the elements out of which the universe was constructed''}
(from Bulfinch \cite{Bulfinch}).
Plato's (c. 427- c. 347 B.C.) emphasis in geometry, in particular his dictum
{\it ``God geometrizes''} \footnote{ In {\it Convivialium disputationum, liber 8,2},
Plutarch stated, {\it ``Plato said God geometrizes continually.''}}
was interpreted by Gauss (1777-1855) as
{\it ``o theos arithmetizei,''} or {\it ``God computes.''}
The vision of a clockwork universe is probably best characterized
by the (probably apocryphal) story, that when Laplace was asked by Napoleon
how God fitted into his secular system of {\it M\'ecanique C\'eleste}, he replied
\cite[p. 538]{boyer},
{\it ``I have no need for that hypothesis''}
\footnote{
In his memoires written on St. H\'el\`ene, Napoleon states that
he removed Laplace from office as Minister of the Interior
\cite[p. 536]{boyer} after only six weeks
{\it ``because he brought the spirit of the infinitely small into the government.''}
}.

In his famous
lecture delivered before the International Congress of Mathematicians at Paris in 1900,
Hilbert (1862-1943) enumerated
twenty-three problems, among them
the compatibility of the arithmetical axioms  (\# 2),
the mathematical treatment of the axioms of physics  (\# 6), and the
determination of the solvability of a diophantine equation  (\# 10)
\footnote{
{\tt http://babbage.clarku.edu/}$\widetilde{\;\;}\,${\tt djoyce/hilbert/problems.html}
}.
G\"odel (1906-78), as well as Turing (1912-54)
contributed towards the (negative) solution of \# 2 \& \# 10.
They pursued a formalization of mathematics by coding of axiomatic systems,
either by the uniqueness of prime factorization
or by their representation as (universal) computer programs
\footnote{
In a postscript dated from June 3rd, 1964 \cite[p. 369-370]{godel-ges1},
G\"odel's opinion is clearly expressed,
{\em ``$\ldots $ due to A. M. Turing's work [[on the universal Turing machine]],
a precise and unquestionably adequate definition of the general
concept of formal system can now be given,
the existence of undecidable arithmetical propositions and
the non-demonstrability of the consistency of a
system in the same system can now be proved rigorously
for every consistent formal system containing
a certain amount of finitary number theory.
$\ldots $
Turing's work gives an analysis of the concept of
``mechanical procedure''
(alias ``algorithm'' or ``computation procedure'' or ``finite combinatorial procedure'').
This concept is shown to be equivalent with that of a ``Turing machine.''
A formal system can simply be defined to be any
mechanical procedure for producing formulas, called provable formulas.''}
}.

For the first time in human history,
we are able to articulate precisely what we mean when
discussing computations.
Turing's universal computer model is modelled after the syntax of
everyday pencil and paper operations which children learn at school.
The paper lines are unwound into a tape, and whatever rules there are for computing
can be represented by the combination of tape, finite memory and simple read-write operations
of the Turing machine.

The notion of universal computation is {\it robust}
in the sense that any universal computer can emulate
any other universal computer
(regardless of efficiency and overhead),
so that it does not really matter which one is
actually implemented.
In a sense, the entire class  of universal computer
 counts as a single computer,
because they are all equivalent with
respect to algorithmic emulation of one another.

Robustness is a very important concept for the matter of computational universes, because
it is not really important on which particular models or
hardware these universes are implemented; they are all in the same equivalence class.
Apart from the translation from one coding scheme to another,
each one of them is equivalent to the entire class.
So, when it comes to their generic properties, it is not really important
whether automaton universes are modelled to be Cellular Automata,
Turing Machines, colliding billiard balls \cite{fred-tof-82},
or biological substrates.
All of this  means that one is free to choose whatever
computational model suits best the particular purpose one has in mind.

G\"odel, Tarski, Turing and Chaitin, among others, revealed that,
stated pointedly, mathematical ``truth'' extends formal ``provability.''
Mathematics is incomplete, and
there always will be
true theorems about a particular formal system of axioms
(sufficiently rich to contain arithmetic), such as consistency,
which are not provable ``from within'' that system
\footnote{
G\"odel's own thoughts on the interpretation of his results are formulated very nicely in a
reply to a letter by A. W. Burks,
reprinted in \cite[p. 55]{v-neumann-66},
{\it ``I think the theorem of mine which von Neumann refers
to is $\ldots$ the fact that a complete epistemological description of a
language A cannot be given in the same language A,
because the concept of truth of sentences of A cannot be defined in A.
It is this theorem which is the true reason for the existence of
undecidable propositions in the formal systems containing arithmetic.
I did not, however, formulate it explicitly in my paper of 1931
but only in my Princeton lectures of 1934.
The same theorem was proved by Tarski in his paper on the concept of truth
$\ldots$''}
}.
%Early attempts by popper

Wigner considered
{\it "the unreasonable effectiveness of mathematics in the natural sciences"}
\cite{wigner},
which is usually taken for granted but which,
upon inspection, seems unfounded.
One obvious solution to this bewilderment seems to be the
Pythagorean assumption that numbers are
the elements out of which the universe was constructed;
and what appears to us as the laws of Nature are just
mathematical theorems or computations.
Notice that, whereas G\"odel once and for all settled the question of a complete
finite description of mathematics to the negative,
the question of whether or not a finite  mathematical treatment of the axioms of physics
exists (Hilbert's problem \# 6) remains open.

Another thread was opened by
Edward Moore.
Puzzled by the quantum mechanical feature of complementarity,
Moore
conceived a finite deterministic model of
complementarity capable of being run on a computer \cite{e-f-moore,conway}.
This formalization of complementarity, not in terms of Hilbert space quantum mechanics,
but by constructive algebraic, even finitistic, means,
may be perceived as the continuation of the Turing
program to formalize the  notion of ``algorithm'' or ``computation''
by conceptualizing it as a concrete machine model.

In another development,
Von Neumann (preceded by Ulam \cite{Ulam-50})
constructed a two-dimensional cellular array of finite deterministic automata which
are connected to their neighbors
such that the state of each one of these automata is determined by the previous states
of itself and of its neigbors \cite{v-neumann-66}.
He was able to show that such cellular automata (CA) could not only be in the
robust class of universal computers, but that entities inside such arrays could
reproduce themselves by holding their own descriptional code and the algorithmic means
to construct identical copies of themselves.

Stimulated by Von Neumann's concept of CA, Konrad Zuse,
the creator of one of the first general purpose digital computers,
suggested to look into the idea
that physical space itself might actually be such a ``calculating space''
(``Rechnender Raum'') \cite{zuse-67,zuse-69,zuse-94}.
In this view, the  physical objects exist as computational entities immersed
in such a computational medium.
Zuse became fascinated by the idea of going
beyond quantum mechanics in discretizing physics
\footnote{
Quantum theory just discretizes the number of quanta within a mode,
yet the modes themselves are still continuous.
},
a vision he shared with the late Einstein
\footnote{
In \cite[p. 163]{ein1}, Einstein states,
{\it
``There are good reasons to assume that nature cannot be represented by a continuous field. From quantum theory it could be inferred with certainty that a finite system with finite energy can be completely described by a finite number of (quantum) numbers. This seems not in accordance with continuum theory and has to render trials to describe reality with purely algebraic means. However, nobody has any idea of how one can find the basis of such a theory.''
}
}
and many researchers, among others Fredkin, Toffoli, Margolus, and Wolfram.

Fredkin and Toffoli investigated reversible CA,
in which the global temporal evolution can be inverted uniquely.
That is, any CA configuration has a unique predecessor and a unique successor.
Note that, if the evolution is a bijective map;
i.e., is one-to-one for every single cell,
then the global array is a reversible CA as well.
(The converse need not be satisfied.)
For a concise account \footnote{
Reversibility of CA should  not be confused with Bennett's strategy
to produce ``reversible'' calculations from irreversible ones
by  temporarily copying their intermediate results and permanently
copying their final result,
thereby setting the computing agent to its initial state,
as well as retaining
the result of the computation \cite{bennett-73}.
Any such operation must necessarily allow for copying; i.e., for a one-to-many evolution,
which is clearly not meant here.
}
the reader is referred to
the reviews by Toffoli \& Margolus
\cite{toffoli-margolus-90},
Fredkin
\cite{fredkin} and Wolfram
\footnote{
Wolfram  recently self-published a long-awaited and widely publicised book
which, among other issues, deals with some of the rules categorized for one-dimensional automata
\cite{wolfram-2002} and their conceivable physical applications.
It has been received ambivalently with reviews ranging from the author doing nice computer graphics
to becoming the biggest physics guru of all times
\cite{gray-2003}.
Wolfram has attracted a lot of attention for this subject;
and it can only be hoped that the many claims made in this {\it opus}
will not deter others.}.
In a reversible world, nothing is lost or gained; and all revelations are
permutated back and forth.
In this sense, the very concept of question and answer, of problem and solution,
of past, present and future, and thus of a directed ``lapse of time,''
remains relative, subjective and conventional
\cite{godel-sch}.

\section{Intrinsic randomness and undecidability}

Contemporary theoretical physics postulates at least three types
of randomness:
(i) the ``chaotic'' randomness residing in the initial conditions,
which are assumed to be ``drawn'' ({\it via} the postulated axiom of choice)
from a ``continuum urn.''
Almost all elements of the continuum are
nonrecursively enumerable and even random; i.e., algorithmically incompressible
\cite{chaitin3,chaitin:01,calude:02};
(ii) the random occurrence of individual quantum events such as a detector click;
(iii) complementarity; i.e., the impossibility to measure two or more
observables with arbitrary precision at once.

\subsection{Computational complementarity}
As already mentioned, Moore \cite{e-f-moore} invented (parts of) finite automata theory
to formalize and model physical complementarity.
Research in this area became totally separated from its original physical perspective
and developed into a beautiful algebraic theory of its own \cite{brauer-84}.
Finkelstein \cite{finkelstein-83} rediscovered Moore's paper
and coined the term ``computational complementarity.''
Its concrete logico-algebraic structure has been investigated by the author
in a series of papers with  C. and E. Calude, Khoussainov, Lipponen, Schaller and others
\cite{svozil-93,schaller-96,cal-sv-yu,svozil-ql},
also in the context of reversible computation \cite{sv-aut-rev,svozil-2002-kyoto}.
Automaton logic turns out to be logically equivalent
\cite{svozil-2001-eua}
to generalized urn models
\cite{wright:pent,wright},
indicating that the associated logico-algebraic structure
is more robust than could be assumed from those single model types alone.

Arguably, the simplest automaton model featuring complementarity
is a finite (Mealy) automaton in which the sets contain
three internal states
$S=\{1,2,3\}$,
three input symbols
$I=\{1,2,3\}$,
and two output symbols
$O=\{0,1\}$.
Let, for $s\in S$, $i\in I$,
the (irreversible ``guessing'') output function be
$\lambda (s,i)=\delta_{si}$.
The (irreversible) transition function just steers the automaton
into a state corresponding to the input symbol; i.e.,
$t(s,i)=i$.
The problem of finding an unknown initial state by
analysis of experimental input--output sequences
yields a partitioning of the internal states
$\{\{1\},\{2,3\}\}$,
$\{\{1,3\},\{2\}\}$, and
$\{\{1,2\},\{3\}\}$,
according to the input $1$, $2$, and $3$, respectively.
Every one of the partitions constitutes a Boolean algebra
whose elements are comeasurable.
The pasting of these three Boolean algebras along their common elements
(in this case just $\emptyset$ and $\{1,2,3\}$) yields a
nonclassical, nondistributive logical structure $MO_3$, which is also
realized by the algebra
of propositions associated with the electron spin state measurements along three
different directions.

A systematic investigation shows that the logico-algebraic structures
arising from computational complementarity are very similar to those encountered in
quantum logics \cite[Sec. 3.5.2]{svozil-ql}. In particular,
any finite quantum (sub-)algebra can be represented as an automaton logic and thus
can be modelled with a finite automaton.
Clearly, infinite quantum structures, such as the continuous ``Chinese lantern'' lattices $MO_c$
involved in electron spin state measurements in continuous directions,
or quantum contextuality, cannot be modelled with a finite automaton.

Reversible finite automata have been introduced
by the author  \cite{sv-aut-rev,svozil-ql,svozil-2002-kyoto}
as Mealy automata whose input and output symbols are identical.
Consider the Cartesian product $S\times I$ of the set of automaton states $S$
with the set of input symbols $I$, arranged in vector form $SI=((s_1,i_1),\ldots )$;
as well as the Cartesian product $S\times O$ of the set of automaton states
with the set of output symbols $O$, again arranged in vector form $SO=((s_1,o_1),\ldots )$.
The transition and output functions and thus the automaton computation can then be
formalized by a matrix multiplication $SO = SI\cdot P$, where $P$ is the matrix
associated with the combined transition and output function  $P: SI\rightarrow SO$.
Reversibility implies that these matrices $P$ are permutation matrices
(i.e., every row and every column contains exactly one entry ``1,''
all other entries are zeroes).

The most general probabilistic state of all reversible Mealy automata associated with a particular
matrix ``dimension'' can be represented as the weighted convex sum over all permutation
matrices of this dimension. The result is a doubly stochastic matrix (i.e., the sum of the
real components of every row and column adds up to one).
Formally, let $\psi : SI\times SO \rightarrow [0,1]$ be the transition probability.
The convex sum of all transition probabilities is one; i.e., $\sum_{SI,SO}\psi (SI,SO)=1$.

A  modification of this model, according to Fortnov \cite{fortnov-03,be-va},
captures the class {\bf BQP}, the class of efficiently
quantum computable problems. The modification is twofold:
first, the weighted sum over all permutation matrices contains
coefficients $\psi$, called ``probability amplitudes,'' which take on arbitrary rational values
including {\em negative} values.
Secondly, in order for the ``quantum'' probabilities to be positive,
the probability amplitudes $\psi$  have to be squared.
These two modifications---negativity and square values---mark
a demarcation line between quantum and classical computation.

Although computational complementarity will not be reviewed any further here,
it should be mentioned that Moore conceptualized
input/output experiments on finite automata,
making a distinction between ``intrinsic'' cases where only one automaton is available,
and those in which an arbitrary number of identical copies are accessible.
In the latter case, there is no complementarity, because after any experiment
it is always possible to dispose of the used automaton and get a fresh
automaton copy for further experiment(s).

The intrinsic, embedded, case is the one experienced in physics,
because the observer cannot escape and always is part of the
(``Cartesian prison'' \cite[Meditation 5,15]{descartes-meditation}) system.
Due to restrictions in copying and cloning, it is not possible, for
instance, to obtain an identical copy of a single photon or electron in a nonclassical state.
And only in the single automaton case there is a chance to experience complementarity,
for only in this case it may happen that, after answering to some
query, the automaton undergoes an irreversible transition,
making it impossible for the experimenter to probe a different observable (and {\it vice versa}).
Reversibility does not change the picture, since if both the
observer and the observed object were immersed in a reversible environment,
then the experiment could be ``undone'' and
the original automaton state reconstructed only at the price of loosing
all the information gathered so far \cite{svozil-2000interface}.
This is an analogue to the quantum eraser experiment \cite{hkwz} and
other setups (e.g., \cite{greenberger2}) developed for demonstrating the
feasibility of a
reconstruction of quantum states.

Bear in mind that complementarity is not only a feature of exotic finite models which were
specially crafted for this particular purpose.
Since these finite models represent a subset of objects that  can be
simulated by any universal computer, such as a CA or a Turing machine,
complementarity is, in a sense, a generic
and robust property of all computational universes.

\subsection{Intrinsic undecidability}

The quest to translate G\"odel-Turing type recursion theoretic undecidability into physics
has a long history.
G\"odel himself did not believe that his results have any relevance for physics,
at least not for quantum physics
\footnote{In
 \cite[140-141]{bernstein}, Bernstein writes,
{\it Wheeler said, ``I went to G\"odel, and I asked him,
`Prof. G\"odel, what connection do you see between your incompleteness
theorem and Heisenberg's uncertainty principle?'
I believe that Wheeler exaggerated a little bit now.
He said, `And G\"odel got angry and threw me out of his office!'
Wheeler blamed Einstein for this. He said that Einstein had
brain-washed G\"odel against quantum mechanics and against Heisenberg's uncertainty principle!}
(The author has asked professor Wheeler and got this anecdote confirmed.)
}.
Early on,  Popper speculated about limits of forecast in the light
of these findings \cite{popper-50}.
More recent undecidability results are based on physical configurations
which are provably unsolvable through the reduction
to the halting problem (e.g., \cite{komar,kanter,moore}).

Indeed, since the Turing machine is modelled after a paper and pencil real world scenario,
universal computers can be embedded into certain physical
systems capable of universal computation.
Undecidabilities can then be obtained almost as a ``free lunch;''
i.e.,  by reduction to the recursive unsolvability
of certain prediction problems, such as the halting problem.

So, why do people such as Casti
\footnote{
Casti (co-)organized two conferences; one in Santa Fe \cite{casti:94-onlimits}
and one in Abisko  \cite{casti:96-onlimits},
bringing together many who were interested in this issue at the time.
}, who had  been very interested
in the subject, consider this issue as a ``red herring?''
Maybe because so far not a single problem of
relevance in physics not constructed for this particular purpose
is provably undecidable.

\subsection{Continuum {\it versus} discrete physics}

The conceptualization of the number
system---from just a few finger counts to the natural numbers,
the integers, rationals,
reals \cite{drobot} and further on to complex numbers, quaternions and hyperreals
is undoubtedly one of the most beautiful and greatest achievements of humanity.
Nevertheless, as more and more abstractions enter these great patterns of thought,
one is compelled to question
their practical physical relevance.
Clearly, for instance, infinite divisibility (from the rational onwards)
and continuity (from the reals onwards)
find strong pragmatic justifications by their
applicability to almost all branches of theoretical physics,
including quantum mechanics.
Even so,  some doubts as to the appropriateness
of transfinite concepts in physical modelling
remain \cite{bridgman}.
Let us state the following correspondence principle
between physical phenomena and their models \cite{svozil-set}:
{\em every feature of a computational model should be reflected by the
capacity of the corresponding physical system.
Conversely, every physical capacity, in particular of a physical theory,
should correspond to a feature of an appropriate
computational model.}

Nature does not seem to allow Zeno squeezing
\cite{weyl:49,gruenbaum:74,thom:54,benna:62,pit:90,hogarth1,ear-nor:93,svozil-93,sv-aut-rev}
and other transfinite processes.
It could therefore be conjectured that, as
physical systems do not possess adequate transfinite capacities,
only finite computational models ought to be acceptable for theoretical modelling.
This still admits  universal computation and finite automata,
but it wipes out classical, nonconstructive continua.

Having said this, there may be some indication of absolute randomness
involved in certain quantum measurements, though.
Suppose a single electron is prepared in a particular spin state in
one direction $\theta_p$.
Assume further that its spin state is not measured in this particular direction,
but in another direction $\theta_m$.
Then quantum mechanics predicts that the probability that identical
spin states are measured is $\cos^2[(\theta_p-\theta_m)/2]$; for a
non-identical result the probability is $1-\cos^2[(\theta_p-\theta_m)/2]=\sin^2[(\theta_p-\theta_m)/2]$
(classically, one would expect linear dependencies on the measurement angles,
such as $1- \vert \theta_p-\theta_m\vert /\pi$
and  $\vert \theta_p-\theta_m\vert /\pi $, respectively).
Moreover, quantum mechanics postulates
that these outcomes are  stochastic  and cannot be reduced
to some form of microscopic law governing the single measurement outcomes.
At $\theta_p-\theta_m = \pi /2$, a series of such experiments,
when coded into a $0,1$-sequence, is postulated by the quantum mechanical canon
to render an algorithmically incompressible random sequence
\cite{svozil-qct}; a fact
which can be used to construct a plug-in device \cite{zeilinger:qct};
just like another card which can be inserted into a computer and
facilitates the desired function, in this case the production of
random data.

Here seems to be a physical source of absolute randomness
\cite{chaitin3,chaitin:01,calude:02} which appears almost totally ``free''
of any computational costs.
Just detune preparation and measurement
to attain the goal of a perfect random number generator.
Indeed, this quantum postulate of microphysical randomness
seems to be a remarkable fact,
in particular since randomness is a valuable resource
which, in the context of universal computation, cannot be obtained ``for nothing.''
Although ``almost all'' reals are random, it is hard (indeed impossible)
to come by any concrete element with that property.
The closest one could get may be Chaitin's $\Omega$ number,
which is the Kraft sum of the length of all prefix-free halting programs
on some universal computer (see \cite{chaitin3,chaitin:01,calude:02} for details).
It is even possible to write down a finite
program for computing the first bits of $\Omega$,
but for better precision
there is no computable radius of convergence certifying that
a particular finite sequence is the starting sequence for $\Omega$.
In that respect $\Omega$ resembles Specker's sequence of rational numbers
 with non-recursive limit,
or the Busy Beaver function \cite{rado,brady,chaitin-bb}.

So, is every electron a point particle capable of transfinite computations?
While electrons do not seem to possess any capacity of universal computation at all,
they appear just to be perfect random number generators.
That is indeed amazing!
Maybe we just have not listened carefully enough when crafting the
computational models appropriate for physics.
Is Turin's universal computer model, appended with an additional
``random oracle'' plug-in, sufficient?

In another scenario, an electron might just be coded to carry the answer
to a single question; e.g., related to its spin state in a particular
direction. If requested to answer  a different question,
such as about its spin state in a different direction, it might just
churn out random nonsense \cite{zeil-99}
according to Malus' law  \cite{zeil-bruk-99,zeil-bruk-02}.
Or, it may need an interface, an environment or measurement apparatus
translating the observer's question to the language (or question)
understandable by the object \cite{svozil-2000interface,svozil-2002-kyoto},
thereby introducing stochastic noise by uncontrollable macroscopic processes.
So far, these are all metaphysical speculations which need to be sorted out by
operational means, i.e., by experiment.

\subsection{Nonlocality \& contextuality}

Quantum nonlocality is a phenomenon which can be quite easily described,
yet remains mysterious.
Consider again the spin state measurements of electrons. Let us assume that
it is possible to produce  two particles in a singlet state, such that,
when their spin is measured along an arbitrary but identical direction,
their spin states are opposite.
Now, consider the correlation of their spin states when measured along
arbitrary but different directions.
As it turns out, if the directions are different from $0$ and
from integer multiples of
$\pi /2$ or $\pi$, the quantum correlations are either weaker
or stronger than the classical
correlations.
This is related to the difference of the aforementioned quantum probabilities
{\it versus}
the classical ones.
In terms of elementary physical events,
one obtains more or fewer joint clicks in the detectors
measuring the spin states
than would be conceivable classically for any such state.
The doctrine of ``peaceful coexistence'' between relativity theory and quantum mechanics
\cite{shimony2}
assures that this feature cannot be used for
faster than light quantum signalling
\cite{herbert,wo-zu,glauber}.

Can CA with local neighbourhood cell evolution reproduce quantum-type nonlocality?
That seems to be a hard problem, in particular if one
clings to the idea of evolution functions which only depend on the neighbourhood,
a property which surely seems to be a constituent element in the definition of CA.
Indeed, with regards to nonlocality, little convincing evidence and
comfort has been given so far by the CA community.
Zuse mentions the chess metaphor of the bishop, a piece which can move in single-colour diagonal direction
only, thereby exerting a nonlocal influence on the entire  chessboard
\cite{zuse-94}.
But how could the entire   chessboard know
of the bishop's motion if information
can only propagate by one cell per time step?
Considering quantized cells is no solution,
because the quantum nonlocalities introduced by
proper normalization of the entire ray wave function
comes as no surprise
\cite{fu-groe-93}:
quantum behaviour of a quantized system
is indeed to be expected.

Another, entirely different and radical possibility would be
to give up the notion of ``calculating space'' and consider a
computational substratum which is nonlocal from the very beginning.
In this approach, the cellular space does not correspond to anything
which is spatially extended from a physical point of view,
such as the  tesselated configuration space  Zuse had in mind.
Rather, it might be some kind of generalized phase space, in which
physical states are discrete.
This resembles the ``old'' quantum mechanics
of Planck and Einstein
\footnote{
As expressed by Planck \cite[p. 387]{planck:1916},
{\it ``Again it is confirmed that the quantum hypothesis
is not based on energy elements but on action elements,
according to the fact that the volume of
phase space has the dimension $h^f$.''}
} and known as Bohr-Sommerfeld ``quantization.''

Contextuality is another controversial issue which
is discussed in the quantum context \cite{redhead}.
One may argue that as it cannot be operationalized anyway,
contextuality is a property of almost pure theoretical value,
such as counterfactuals or scholastic {\it infuturabilities.}
In continuum theory, there are ``exotic'' ways to come by
\cite{pitowsky-82,meyer:99},
but this is no option for a discrete model.
At first sight, classical computers seem to be value definite
and noncontextual, but a closer inspection reveals
that there are subtleties to be kept in mind.
Value definiteness need not imply that an
agent is prepared to answer {\em any} experimental
question.
Indeed, in contradistinction to Kant's
transcendental ideal
\footnote{
In the {\it 3. Hauptst\"uck, 3. Abschnitt.
Von dem transzendentalen Ideal (Prototypon transscendentale)},
of {\it ``Kritik der reinen Vernunft,''} \cite{kant-kdrv}, Kant states,
{\it ``But again, everything, as regards its possibility, is also subject
to the principle of complete determination, according to which one
of all the possible contradictory predicates of things must belong
to it.''}
The German original reads, {\it ``Ein jedes Ding aber, seiner M\"oglichkeit nach,
steht noch unter dem Grundsatze der durchg\"angigen Bestimmung,
nach welchem ihm von allen m\"oglichen Pr\"adikaten der Dinge,
sofern sie mit ihren Gegenteilen verglichen werden, eines zukommen mu\ss.''}
},
and scholastic, theological speculations whether or not
the omniscience of God extends to events which
would have occurred if something  had happened that did not
happen, which have been so powerfully
formalized into a finitistic proof
(cf.
\cite[p. 243]{specker-60} and
\cite[p. 179]{specker-ges}),
some properties may not be properly definable for certain computational agents,
and therefore may not be operational.
For example, if an agent trained to wash dishes
is confronted with the task to write a book on hiking trails
in New Zealand's Waitakere ranges, it
will most certainly be at a complete loss.
%(Sometimes, the converse is true as well.)
Or an agent advised to direct some parties
to a path on the right hand side when asked
for right or left, will most certainly be at a
loss when asked whether to proceed up or down.
The agent simply would not be programmed and thus not be prepared to answer
any type of question,
but rather only a small selection from among all conceivable ones.

\section{Intrinsic, embedded observer mode}
Computational complementarity and undecidability in general
are good examples of how the science of systems may enter physics.
Unless one accepts the concept of an ``intrinsic embedded mode,''
computational complementarity disappears into thin air.
And since system science seems foreign to most physicists,
it is hard to see if and when such concepts will be more
broadly recognized.

As with all general concepts, it is hard to pinpoint
when exactly the concept of an intrinsic embedded observer
was formulated for the first time
\footnote{
One is also tempted to mention
Archimedes' {\it ``points outside the world from which one could move the earth.''}
Mind that Archimedes' use of ``points outside the world''
was in a mechanical rather than in a metatheoretical context:
he claimed to be able to move any given weight by any given force, however small.
}.
Boskovich
\cite{bos} around 1755 referred to the fact that embedded
observers cannot recognize an overall change (squeeze, dilatation and contraction) of the system size
\footnote{
In \cite{bos}, Boskovich states,
{\it ``$\ldots$ And we would have the same impressions if,
under conservation of distances, all directions would be rotated by the same angle, $\ldots$
And even if the distances themselves would be decreased,
 whereby the angles and the proportions would be conserved, $\ldots$:
even then we [[the observers]] would have no changes in our impressions. $\ldots$
A movement, which is common to us [[the observers]] and to the Universe,
cannot be observed by us; not even if everything would be stretched or shrinked by an arbitrary amount.''
}
}.
More recently, Toffoli \cite{toffoli:79} discussed the role of the observer in uniform systems.
Embedded observers are {\em the} big issue in relativity
theory, because Einstein insisted on operational
methods available within the system only
in defining clocks, length scales,
and when comparing them
\footnote{
The author had some problem to publish a paper on embedded observers in
relativity theory, apparently because of the rather unconventional nature of
the subject.  However, after an appeal, the paper became preprint \#
LBL-16097 \cite{svo-83} and was later published in a revised version \cite{svo5} (see also
\cite{svo-86}).  I write this to encourage young researchers not to give up in
pursuing their own nonfashionable ideas  \cite{dyson-unfash}.
%The author had some problem to publish a paper on embedded observers in
%relativity theory
%an LBL preprint from Lawrence Berkeley National Laboratory.
%The following story is a recollection from my memory.
%The request to publish my manuscript
%as an LBL preprint
%got almost censured by the head of the theory division Geoffrey Chew on the basis of
%an evaluation by Henry Stapp, who  disliked it.
%Like Chew, Stapp came from
%bootstrapping theory of nuclear forces, which had been outdated even then.
%He had started doing research in the foundations
%of quantum physics. Besides, he had just published a speculative
%attempt to relate psychic phenomena to quantum
%physics in the daily newspaper {\it San Francisco Examiner,}
%a suggestion which was rather sceptically received by most of the LBL physics community.
%I appealed to the head of the physics department, Professor Jackson
%(who had written the classical theoretical physics bestseller {\it Classical Electrodynamics}
%\cite{jackson}), who overruled
%Stapp's verdict and Chew's decision.
%The paper became  preprint \# LBL-16097  \cite{svo-83}
%and was later published in a revised version  \cite{svo5} (see also \cite{svo-86}).
%I am not writing this to whine or accuse any fellow
%researcher---in fact, I like Chew and Stapp personally,
%and given the rather unconventional
%nature of the subject, one can hardly blame them for their
%reluctance---but to encourage young researchers
%not to give up in pursuing own nonfashionable ideas \cite{dyson-unfash}.
}.
R\"ossler  \cite{roessler-87,roessler-92} and the author
\cite{svo-83,svo5,svo-86,svozil-94,svozil-93}, independently
share similar concepts,
although R\"ossler's emphasis has been on the
role of the interface between observer and observed object
\cite{roessler-98}
rather than on concrete examples of automaton logic or space-time frames
\footnote{
When I was invited to participate to a Linz {\it Ars Electronica}
 conference on ``Endophysics'' in 1992,
I was almost shocked by the similarity between
R\"ossler's thoughts and the ones I had pursued.
%so great difficulties making somebody understandable.
One follow-up meeting was organized by Atmanspacher in Germany \cite{atman:93}.
I later learned that for researchers trained in mathematical system science,
like John Casti, embedded observers sounded like a very familiar, almost self-evident concept.
\cite{casti:90,casti:92,casti:92b}.
}.

\section{Space time frames of intrinsic observers}

Relativity theory has altered the way we think of space
and time  from a formal point of view,  but
the perception of space and time at large,
and what meaning is ascribed to these notions,
has not changed too much:
while pre-relativistic ``Galilean'' type thinking considered
space and time as absolute and immutable, nowadays this role
is ascribed to the relativistic forms of space time coordinates
and their transformation laws.
It is almost as if the attitude of the protagonists remained the same, but their
message changed slightly.

Relativity theory, as introduced by Einstein,
at least in the first, kinematic, part of the seminal 1905 paper \cite{ein-05},
is conceived as a strictly operational theory for embedded, intrinsic
observers. Those observers are bound to use the methods and capabilities of the
system of which they are an integral part;
and they cannot resort to an extrinsic, ``God's eye'' overview of it.
But operationalism is not enough to create space-time frames.
What is also needed (but seldom mentioned although implicitly assumed)
are conventions for measuring time and space,
and for comparing those scales at different locations and different times
in co-moving and other experimental configurations.
Indeed, the International System of units
outrightly declares a previously experimental physical fact
to be convention.
The speed of light is assumed to be constant for all reference frames.
With the mild side assumption of the one-to-oneness (invertibility) of
space-time transformation,
this convention declares the preservation of light cones,
and thus, by the preservation of set theoretic intersections of light cones
such as time- space- and lightlike onedimensional subspaces,
results in affinity and linearity of the transformation laws.
From this point of view, the Lorentz transformation
is a geometric, not a physical entity.
In geometry, this is known as Alexandrov's theorem
\cite{alex1,alex2,alex3,alex-col,borchers-heger,benz,lester}.

So, in a sense, this is  the big picture.
If one requires invariance of some ``fundamental'' speed and bijectivity of the
transformation laws, then the Lorentz-type transformation laws
containing that ``fundamental'' speed
follow.
Thereby, it makes no difference whether the associated observers are embedded in
the ``real'' Universe, in a CA, or in a plum pudding; as long as
these conditions and conventions are met, then
Alexandrov's theorem certifies that the geometry is a relativistic one.
For discrete models, these results will always be only approximations
which are valid down to scales where the discreteness becomes important.

Where is all the physics gone? The answer to this question is that
the physics is in the invariance with respect to any such Lorentz-type transformations.
For example, clocks governed by electromagnetic phenomena will be showing the ``right''
time in all frames if the ``fundamental'' speed is chosen to be the speed of light.
Sound clocks tick invariantly in the respective system if
the ``fundamental'' speed is the speed of sound.
Scales are invariant if the forces stabilizing that scales are electromagnetic ones
and the  ``fundamental'' speed is again chosen to be the speed of light.
So, with these new conventions, the invariance of certain length \cite{peres-84}
and time scales, corresponding to the relativistic form invariance of the
laws governing them, becomes a physical statement
 \cite{bell-92,peres-84,svozil-relrel,svozil-2001-convention}.

The following is an  example \cite{svozil-1996-time} of an
Einstein synchronisation by clocks generating radar coordinates in a
one-dimensional CA with the following evolution rules.
{\scriptsize
\begin{equation}
\begin{array}{lllll}
\varphi ( > ,\_ ,X)  \rightarrow   > ,&
  \varphi (X,\_ , < )  \rightarrow   < ,&
  \varphi (\_ ,\_ ,\_ )  \rightarrow  \_ ,&
  \varphi (X,\_ , > )  \rightarrow  \_ ,&
  \varphi ( < ,\_ ,X)  \rightarrow  \_ ,\\
  \varphi (\_ , > ,\_ )  \rightarrow  \_ ,&
  \varphi (\_ , < ,\_ )  \rightarrow  \_ ,&
  \varphi (\_ , > ,I)  \rightarrow   < ,&
  \varphi (I, < ,\_ )  \rightarrow   > ,&
  \varphi ( > ,I,X)  \rightarrow   * ,\\
  \varphi ( < , * ,X)  \rightarrow  I,&
  \varphi (X, < , * )  \rightarrow  \_ ,&
  \varphi ( * , 1 ,X)  \rightarrow   2 ,&
  \varphi ( * , 2 ,X)  \rightarrow   3 ,&
  \varphi ( * , 3 ,X)  \rightarrow   4 ,\\
  \varphi ( * , 4 ,X)  \rightarrow   5 ,&
  \varphi ( * , 5 ,X)  \rightarrow   6 ,&
  \varphi ( * , 6 ,X)  \rightarrow   7 ,&
  \varphi ( * , 7 ,X)  \rightarrow   8 ,&
  \varphi ( * , 8 ,X)  \rightarrow   9 ,\\
  \varphi ( * , 9 ,X)  \rightarrow   0 ,&
  \varphi ( * , 0 ,X)  \rightarrow   1 ,&
  \varphi ( 0 ,\_ ,X)  \rightarrow  \_ ,&
  \varphi ( 1 ,\_ ,X)  \rightarrow  \_ ,&
  \varphi ( 2 ,\_ ,X)  \rightarrow  \_ ,\\
  \varphi ( 3 ,\_ ,X)  \rightarrow  \_ ,&
  \varphi ( 4 ,\_ ,X)  \rightarrow  \_ ,&
  \varphi ( 5 ,\_ ,X)  \rightarrow  \_ ,&
  \varphi ( 6 ,\_ ,X)  \rightarrow  \_ ,&
  \varphi ( 7 ,\_ ,X)  \rightarrow  \_ ,\\
  \varphi ( 8 ,\_ ,X)  \rightarrow  \_ ,&
  \varphi ( 9 ,\_ ,X)  \rightarrow  \_ ,&
  \varphi (X, * , 0 )  \rightarrow   * ,&
  \varphi (X, * , 1 )  \rightarrow   * ,&
  \varphi (X, * , 2 )  \rightarrow   * ,\\
  \varphi (X, * , 3 )  \rightarrow   * ,&
  \varphi (X, * , 4 )  \rightarrow   * ,&
  \varphi (X, * , 5 )  \rightarrow   * ,&
  \varphi (X, * , 6 )  \rightarrow   * ,&
  \varphi (X, * , 7 )  \rightarrow   * ,\\
  \varphi (X, * , 8 )  \rightarrow   * ,&
  \varphi (X, * , 9 )  \rightarrow   * ,&
  \varphi (X, 1 ,X)  \rightarrow   1 ,&
  \varphi (X, 2 ,X)  \rightarrow   2 ,&
  \varphi (X, 3 ,X)  \rightarrow   3 ,\\
  \varphi (X, 4 ,X)  \rightarrow   4 ,&
  \varphi (X, 5 ,X)  \rightarrow   5 ,&
  \varphi (X, 6 ,X)  \rightarrow   6 ,&
  \varphi (X, 7 ,X)  \rightarrow   7 ,&
  \varphi (X, 8 ,X)  \rightarrow   8 ,\\
  \varphi (X, 9 ,X)  \rightarrow   9 ,&
  \varphi (X, 0 ,X)  \rightarrow   0 ,&
  \varphi (X,I, 0 )  \rightarrow  I,&
  \varphi (X,I, 1 )  \rightarrow  I,&
  \varphi (X,I, 2 )  \rightarrow  I,\\
  \varphi (X,I, 3 )  \rightarrow  I,&
  \varphi (X,I, 4 )  \rightarrow  I,&
  \varphi (X,I, 5 )  \rightarrow  I,&
  \varphi (X,I, 6 )  \rightarrow  I,&
  \varphi (X,I, 7 )  \rightarrow  I,\\
  \varphi (X,I, 8 )  \rightarrow  I,&
  \varphi (X,I, 9 )  \rightarrow  I,&
  \varphi (X,I, 0 )  \rightarrow  I,&
  \varphi (X,I,X)  \rightarrow  I,&
\varphi ( * ,\_ ,X)  \rightarrow  \_ ,\\
\varphi (\_ ,\_ ,I)  \rightarrow  \_ ,&
\varphi (I,\_ ,\_ )  \rightarrow  \_ ,&
\varphi (I, > ,\_ )  \rightarrow  \_ ,&
\varphi (\_ , < ,I)  \rightarrow  \_  .
\end{array}\nonumber
\end{equation}
}
Here,
$X$ stands for any state except the ones already specified.
These rules look a little bit ``murky,''
but they can be simulated by any universal CA and they serve their purpose
to demonstrate clock synchronization procedures.

 Assume two clocks at two arbitrary points $A$ and $B$
in the CA which are ``of similar  kind.''
 At some arbitrary $A$-time $t_A$ a  ray goes from $A$ to $B$. At $B$
it is instantly (without delay)
 reflected  at $B$-time $t_B$ and reaches $A$ again at
 $A$-time $t_{A'}$. The clocks in $A$ and $B$ are {\em synchronized}
 if
$
 t_B-t_A=t_{A'}-t_B$.
 The two-ways ray velocity is given by
$
 {2 \vert {AB}\vert / (t_{A'}-t_A)}=c
$,
 where $\vert AB\vert $ is the distance between $A$ and $B$.
In Fig. \ref{synchro}(a), an example of synchronization between two clocks
$A$ and $B$
is drawn.

What happens with the intrinsic
synchronization and the space-time
coordinates when observers are considered which are in motion with respect to the
CA? For simplicity, suppose a constant motion of $v$ automaton cells
per time cycle. With these units, the ray speed is $c=1$, and $v\le 1$.
There are numerous ways to simulate sub-ray motion on a CA. In what
follows, the case $v=1/3$ will be studied in such a way that every three
CA time cycles the walls, symbolised by ${\tt I}$, move one cell to the
right.

Notice that two clocks which
are synchronized in a reference frame which is at rest with respect to
the CA medium are {\em not synchronized} in their own co-moving
reference frame.
Consider, as an example, the CA drawn in Fig.
\ref{synchro}(b). (Strictly speaking, the CA rule here depends on
a two-neighbor interaction.)
  For $t_A=1$, $t_B=4$, $t_{A'}=5$, and $4-1\neq 5-4$,
if the first clock is corrected to make up for the different time of ray
flights as in Fig. \ref{synchro}(c),
$t_A=2$, $t_B=4$, $t_{A'}=6$, and $4-2 = 6-4$.
Then, this correction amounts to an asynchronicity of the two
ray clocks with respect to the ``original'' CA medium.
\begin{figure}
{\tiny
 \begin{verbatim}
      clock1  A       B  clock2          clock1  A       B  clock2                   clock1  A       B  clock2

______I>__I0__I>______I__I>__I0______   __I>_I0__I___<___I__I>_I0________________   __I>_I1__I___<___I__I>_I0________________
______I_>_I0__I_>_____I__I_>_I0______   __I_>I0__I__<____I__I_>I0________________   __I_>I1__I__<____I__I_>I0________________
______I__>I0__I__>____I__I__>I0______   __I_<*0__I_<_____I__I_<*0________________   __I_<*1__I_<_____I__I_<*0________________
______I__<*0__I___>___I__I__<*0______   ___I>_I1__I>______I__I>_I1_______________   ___I>_I2__I>______I__I>_I1_______________
______I_<_I1__I____>__I__I_<_I1______   ___I_>I1__I_>_____I__I_>I1_______________   ___I_>I2__I_>_____I__I_>I1_______________
______I<__I1__I_____>_I__I<__I1______   ___I_<*1__I__>____I__I_<*1_______________   ___I_<*2__I__>____I__I_<*1_______________
______I>__I1__I______>I__I>__I1______   ____I>_I2__I__>____I__I>_I2______________   ____I>_I3__I__>____I__I>_I2______________
______I_>_I1__I______<*__I_>_I1______   ____I_>I2__I___>___I__I_>I2______________   ____I_>I3__I___>___I__I_>I2______________
______I__>I1__I_____<_I__I__>I1______   ____I_<*2__I____>__I__I_<*2______________   ____I_<*3__I____>__I__I_<*2______________
______I__<*1__I____<__I__I__<*1______   _____I>_I3__I____>__I__I>_I3_____________   _____I>_I4__I____>__I__I>_I3_____________
______I_<_I2__I___<___I__I_<_I2______   _____I_>I3__I_____>_I__I_>I3_____________   _____I_>I4__I_____>_I__I_>I3_____________
______I<__I2__I__<____I__I<__I2______   _____I_<*3__I______>I__I_<*3_____________   _____I_<*4__I______>I__I_<*3_____________
______I>__I2__I_<_____I__I>__I2______   ______I>_I4__I______>I__I>_I4____________   ______I>_I5__I______>I__I>_I4____________
______I_>_I2__I<______I__I_>_I2______   ______I_>I4__I______<*__I_>I4____________   ______I_>I5__I______<*__I_>I4____________
______I__>I2__I>______I__I__>I2______   ______I_<*4__I_____<_I__I_<*4____________   ______I_<*5__I_____<_I__I_<*4____________
______I__<*2__I_>_____I__I__<*2______   _______I>_I5__I___<___I__I>_I5___________   _______I>_I6__I___<___I__I>_I5___________
______I_<_I3__I__>____I__I_<_I3______   _______I_>I5__I__<____I__I_>I5___________   _______I_>I6__I__<____I__I_>I5___________
______I<__I3__I___>___I__I<__I3______   _______I_<*5__I_<_____I__I_<*5___________   _______I_<*6__I_<_____I__I_<*5___________
______I>__I3__I____>__I__I>__I3______   ________I>_I6__I>______I__I>_I6__________   ________I>_I7__I>______I__I>_I6__________
______I_>_I3__I_____>_I__I_>_I3______   ________I_>I6__I_>_____I__I_>I6__________   ________I_>I7__I_>_____I__I_>I6__________
______I__>I3__I______>I__I__>I3______   ________I_<*6__I__>____I__I_<*6__________   ________I_<*7__I__>____I__I_<*6__________
______I__<*3__I______<*__I__<*3______   _________I>_I7__I__>____I__I>_I7_________   _________I>_I8__I__>____I__I>_I7_________
______I_<_I4__I_____<_I__I_<_I4______   _________I_>I7__I___>___I__I_>I7_________   _________I_>I8__I___>___I__I_>I7_________
______I<__I4__I____<__I__I<__I4______   _________I_<*7__I____>__I__I_<*7_________   _________I_<*8__I____>__I__I_<*7_________
______I>__I4__I___<___I__I>__I4______   __________I>_I8__I____>__I__I>_I8________   __________I>_I9__I____>__I__I>_I8________
______I_>_I4__I__<____I__I_>_I4______   __________I_>I8__I_____>_I__I_>I8________   __________I_>I9__I_____>_I__I_>I8________
______I__>I4__I_<_____I__I__>I4______   __________I_<*8__I______>I__I_<*8________   __________I_<*9__I______>I__I_<*8________
______I__<*4__I<______I__I__<*4______   ___________I>_I9__I______>I__I>_I9_______   ___________I>_I0__I______>I__I>_I9_______
______I_<_I5__I>______I__I_<_I5______   ___________I_>I9__I______<*__I_>I9_______   ___________I_>I0__I______<*__I_>I9_______
______I<__I5__I_>_____I__I<__I5______   ___________I_<*9__I_____<_I__I_<*9_______   ___________I_<*0__I_____<_I__I_<*9_______
______I>__I5__I__>____I__I>__I5______   ____________I>_I0__I___<___I__I>_I0______   ____________I>_I1__I___<___I__I>_I0______
______I_>_I5__I___>___I__I_>_I5______   ____________I_>I0__I__<____I__I_>I0______   ____________I_>I1__I__<____I__I_>I0______
______I__>I5__I____>__I__I__>I5______   ____________I_<*0__I_<_____I__I_<*0______   ____________I_<*1__I_<_____I__I_<*0______
______I__<*5__I_____>_I__I__<*5______   _____________I>_I1__I>______I__I>_I1_____   _____________I>_I2__I>______I__I>_I1_____
______I_<_I6__I______>I__I_<_I6______   _____________I_>I1__I_>_____I__I_>I1_____   _____________I_>I2__I_>_____I__I_>I1_____
______I<__I6__I______<*__I<__I6______   _____________I_<*1__I__>____I__I_<*1_____   _____________I_<*2__I__>____I__I_<*1_____
______I>__I6__I_____<_I__I>__I6______   ______________I>_I2__I__>____I__I>_I2____   ______________I>_I3__I__>____I__I>_I2____
______I_>_I6__I____<__I__I_>_I6______   ______________I_>I2__I___>___I__I_>I2____   ______________I_>I3__I___>___I__I_>I2____
______I__>I6__I___<___I__I__>I6______   ______________I_<*2__I____>__I__I_<*2____   ______________I_<*3__I____>__I__I_<*2____
______I__<*6__I__<____I__I__<*6______   _______________I>_I3__I____>__I__I>_I3___   _______________I>_I4__I____>__I__I>_I3___
______I_<_I7__I_<_____I__I_<_I7______   _______________I_>I3__I_____>_I__I_>I3___   _______________I_>I4__I_____>_I__I_>I3___
\end{verbatim}}
(a) $\qquad \qquad \qquad$ $\qquad \qquad $
(b) $\qquad \qquad \qquad$ $\qquad \qquad $
(c)
\caption{Synchronization by ray exchange (a) in a system as rest with respect to a CA;
(b) ray exchange with synchronization defined by (a);
(c) synchronization in co-moving frame. \label{synchro}}
\end{figure}

\section{Now what?}

Despite all these efforts,  including those of the author presented above,
 the computational approach to understanding  universes has so far resulted in little
or no phenomenological impact;
not to speak of any ``killer application''
which would make the few critics and the many
hesitant researchers listen to the subject.
Large segments of theoretical physics appear to be in the very same position
in other areas such as string theory
or quantum gravity
as well, but  this
is no big comfort.
In search for applications of the idea of
computational universes
let us shortly discuss some of the predictions of the subject and
their possible empirical validation or falsification.

\subsection{New range of phenomena}
With regards to the logical order of propositions,
there may exist phenomena perceivable
by intrinsic, embedded observers which cannot happen according to
quantum mechanics but are realizable by finite automata.
The simplest case is characterized by a Greechie hyperdiagram of triangle form,
with three atoms per edge. Its automaton partition logic is given by
\begin{equation}
\{
\{\{1\},\{2\},\{3,4\}\},
\{\{1\},\{2,4\},\{3\}\},
\{\{1,4\},\{2\},\{3\}\}
\}.
\label{2003-cu-e-nc}
\end{equation}
A corresponding Mealy automaton is
$\langle \{1,2,3,4\},\{1,2,3\},\{1,2,3\},\delta =1 ,\lambda \rangle$, where
$
\lambda (1,1)=
\lambda (3,2)=
\lambda (2,3)=
1
$,
$
\lambda (3,1)=
\lambda (2,2)=
\lambda (1,3)=
2
$, and
$
\lambda (2,1)= \lambda (4,1)=
\lambda (1,2)= \lambda (4,2)=
\lambda (3,3)= \lambda (4,3)=
3
$.

\begin{figure}
\begin{center}
\begin{tabular}{ccc}
%TexCad Options
%\grade{\off}
%\emlines{\off}
%\beziermacro{\off}
%\reduce{\on}
%\snapping{\off}
%\quality{0.20}
%\graddiff{0.01}
%\snapasp{1}
%\zoom{1.00}
\unitlength 1.00mm
\linethickness{0.4pt}
\begin{picture}(41.33,42.00)
%\emline(0.33,4.00)(40.33,4.00)
\put(0.33,4.00){\line(1,0){40.00}}
%\end
\put(20.33,4.00){\circle{2.50}}
\put(40.33,4.00){\circle{2.50}}
\put(0.33,4.00){\circle{2.50}}
%\emline(20.33,38.00)(40.33,4.00)
\multiput(20.33,38.00)(0.12,-0.20){167}{\line(0,-1){0.20}}
%\end
%\emline(0.33,4.00)(20.33,38.00)
\multiput(0.33,4.00)(0.12,0.20){167}{\line(0,1){0.20}}
%\end
\put(20.33,38.00){\circle{2.50}}
\put(0.33,0.00){\makebox(0,0)[ct]{$\{2\}$}}
\put(5.33,23.00){\makebox(0,0)[cc]{$\{3,4\}$}}
\put(33.33,23.00){\makebox(0,0)[lc]{$\{2,4\}$}}
\put(20.33,42.00){\makebox(0,0)[cc]{$\{1\}$}}
\put(20.33,0.00){\makebox(0,0)[ct]{$\{1,4\}$}}
\put(40.33,0.00){\makebox(0,0)[ct]{$\{3\}$}}
%\emline(0.33,4.00)(40.33,3.00)
\put(0.33,4.00){\line(1,0){40.00}}
%\end
\put(10.33,21.00){\circle{2.50}}
\put(30.33,21.00){\circle{2.50}}
\end{picture}
&$\qquad$&
%TexCad Options
%\grade{\off}
%\emlines{\off}
%\beziermacro{\off}
%\reduce{\on}
%\snapping{\off}
%\quality{0.20}
%\graddiff{0.01}
%\snapasp{1}
%\zoom{2.00}
\unitlength 0.750mm
\linethickness{0.4pt}
\begin{picture}(100.00,71.00)
\put(5.00,25.00){\circle*{2.00}}
\put(20.00,25.00){\circle*{2.00}}
\put(35.00,25.00){\circle*{2.00}}
\put(50.00,25.00){\circle*{2.00}}
\put(65.00,25.00){\circle*{2.00}}
\put(80.00,25.00){\circle*{2.00}}
\put(95.00,25.00){\circle*{2.00}}
\put(95.00,45.00){\circle*{0.00}}
\put(95.00,45.00){\circle*{2.00}}
\put(80.00,45.00){\circle*{2.00}}
\put(65.00,45.00){\circle*{2.00}}
\put(50.00,45.00){\circle*{2.00}}
\put(35.00,45.00){\circle*{2.00}}
\put(20.00,45.00){\circle*{2.00}}
\put(5.00,45.00){\circle*{2.00}}
\put(50.00,65.00){\circle*{2.00}}
\put(50.00,5.00){\circle*{2.00}}
%\emline(50.00,5.00)(50.00,25.00)
\put(50.00,5.00){\line(0,1){20.00}}
%\end
%\emline(50.00,25.00)(35.00,45.00)
\multiput(50.00,25.00)(-0.12,0.16){126}{\line(0,1){0.16}}
%\end
%\emline(35.00,45.00)(20.00,25.00)
\multiput(35.00,45.00)(-0.12,-0.16){126}{\line(0,-1){0.16}}
%\end
%\emline(20.00,25.00)(5.00,45.00)
\multiput(20.00,25.00)(-0.12,0.16){126}{\line(0,1){0.16}}
%\end
%\emline(5.00,45.00)(35.00,25.00)
\multiput(5.00,45.00)(0.18,-0.12){167}{\line(1,0){0.18}}
%\end
%\emline(35.00,25.00)(20.00,45.00)
\multiput(35.00,25.00)(-0.12,0.16){126}{\line(0,1){0.16}}
%\end
%\emline(20.00,45.00)(5.00,25.00)
\multiput(20.00,45.00)(-0.12,-0.16){126}{\line(0,-1){0.16}}
%\end
%\emline(5.00,25.00)(35.00,45.00)
\multiput(5.00,25.00)(0.18,0.12){167}{\line(1,0){0.18}}
%\end
%\emline(35.00,45.00)(50.00,65.00)
\multiput(35.00,45.00)(0.12,0.16){126}{\line(0,1){0.16}}
%\end
%\emline(50.00,65.00)(50.00,45.00)
\put(50.00,65.00){\line(0,-1){20.00}}
%\end
%\emline(50.00,45.00)(35.00,25.00)
\multiput(50.00,45.00)(-0.12,-0.16){126}{\line(0,-1){0.16}}
%\end
%\emline(50.00,25.00)(50.00,25.00)
\put(50.00,25.00){\line(0,1){0.00}}
%\end
%\emline(35.00,45.00)(65.00,25.00)
\multiput(35.00,45.00)(0.18,-0.12){167}{\line(1,0){0.18}}
%\end
%\emline(65.00,25.00)(95.00,45.00)
\multiput(65.00,25.00)(0.18,0.12){167}{\line(1,0){0.18}}
%\end
%\emline(95.00,45.00)(80.00,25.00)
\multiput(95.00,45.00)(-0.12,-0.16){126}{\line(0,-1){0.16}}
%\end
%\emline(80.00,25.00)(65.00,45.00)
\multiput(80.00,25.00)(-0.12,0.16){126}{\line(0,1){0.16}}
%\end
%\emline(65.00,45.00)(95.00,25.00)
\multiput(65.00,45.00)(0.18,-0.12){167}{\line(1,0){0.18}}
%\end
%\emline(95.00,25.00)(80.00,45.00)
\multiput(95.00,25.00)(-0.12,0.16){126}{\line(0,1){0.16}}
%\end
%\emline(80.00,45.00)(65.00,25.00)
\multiput(80.00,45.00)(-0.12,-0.16){126}{\line(0,-1){0.16}}
%\end
%\emline(65.00,25.00)(50.00,5.00)
\multiput(65.00,25.00)(-0.12,-0.16){126}{\line(0,-1){0.16}}
%\end
%\emline(50.00,5.00)(35.00,25.00)
\multiput(50.00,5.00)(-0.12,0.16){126}{\line(0,1){0.16}}
%\end
%\put(20.00,25.00){\vector(3,-2){30.00}}
\put(20.00,25.00){\line(3,-2){30.00}}
%\put(50.00,5.00){\vector(3,2){30.00}}
\put(50.00,5.00){\line(3,2){30.00}}
%\emline(95.00,25.00)(50.00,5.00)
\multiput(95.00,25.00)(-0.27,-0.12){167}{\line(-1,0){0.27}}
%\end
%\emline(50.00,5.00)(5.00,25.00)
\multiput(50.00,5.00)(-0.27,0.12){167}{\line(-1,0){0.27}}
%\end
%\emline(5.00,45.00)(50.00,65.00)
\multiput(5.00,45.00)(0.27,0.12){167}{\line(1,0){0.27}}
%\end
%\emline(50.00,65.00)(20.00,45.00)
\multiput(50.00,65.00)(-0.18,-0.12){167}{\line(-1,0){0.18}}
%\end
%\emline(65.00,45.00)(50.00,65.00)
\multiput(65.00,45.00)(-0.12,0.16){126}{\line(0,1){0.16}}
%\end
%\emline(50.00,65.00)(50.00,65.00)
\put(50.00,65.00){\line(0,1){0.00}}
%\end
%\emline(50.00,65.00)(80.00,45.00)
\multiput(50.00,65.00)(0.18,-0.12){167}{\line(1,0){0.18}}
%\end
%\emline(80.00,45.00)(80.00,45.00)
\put(80.00,45.00){\line(0,1){0.00}}
%\end
%\emline(95.00,45.00)(50.00,65.00)
\multiput(95.00,45.00)(-0.27,0.12){167}{\line(-1,0){0.27}}
%\end
%\emline(50.00,65.00)(50.00,65.00)
\put(50.00,65.00){\line(0,1){0.00}}
%\end
%\emline(50.00,45.00)(65.00,25.00)
\multiput(50.00,45.00)(0.12,-0.16){126}{\line(0,-1){0.16}}
%\end
\put(0.00,16.00){\makebox(0,0)[cc]{$\{2\}$}}
\put(16.00,16.00){\makebox(0,0)[cc]{$\{3,4\}$}}
\put(31.00,16.00){\makebox(0,0)[cc]{$\{1\}$}}
\put(47.00,16.00){\makebox(0,0)[cc]{$\{2,4\}$}}
\put(65.00,16.00){\makebox(0,0)[cc]{$\{3\}$}}
\put(98.00,16.00){\makebox(0,0)[cc]{$\{2\}$}}
\put(84.00,16.00){\makebox(0,0)[cc]{$\{1,4\}$}}
\put(50.00,-1.00){\makebox(0,0)[cc]{0}}
\put(50.00,71.00){\makebox(0,0)[cc]{1}}
\put(-1.00,50.00){\makebox(0,0)[cc]{$\{1,3,4\}$}}
\put(16.00,50.00){\makebox(0,0)[cc]{$\{1,2\}$}}
\put(31.00,50.00){\makebox(0,0)[cc]{$\{2,3,4\}$}}
\put(47.00,50.00){\makebox(0,0)[cc]{$\{1,3\}$}}
\put(65.00,50.00){\makebox(0,0)[cc]{$\{1,2,4\}$}}
\put(98.00,50.00){\makebox(0,0)[cc]{$\{1,3,4\}$}}
\put(84.00,50.00){\makebox(0,0)[cc]{$\{2,3\}$}}
\put(50.00,25.00){\line(3,2){30.00}}
\put(50.00,25.00){\line(3,4){15.00}}
\end{picture}
\\
$\;$
\\
a)&&b)
\end{tabular}
\end{center}
\caption{\label{f-gh-tria}
a) Greechie and b) Hasse diagram of a logic featuring complementarity
which is not a quantum logic but which is embeddable in a Boolean logic.
}
\end{figure}
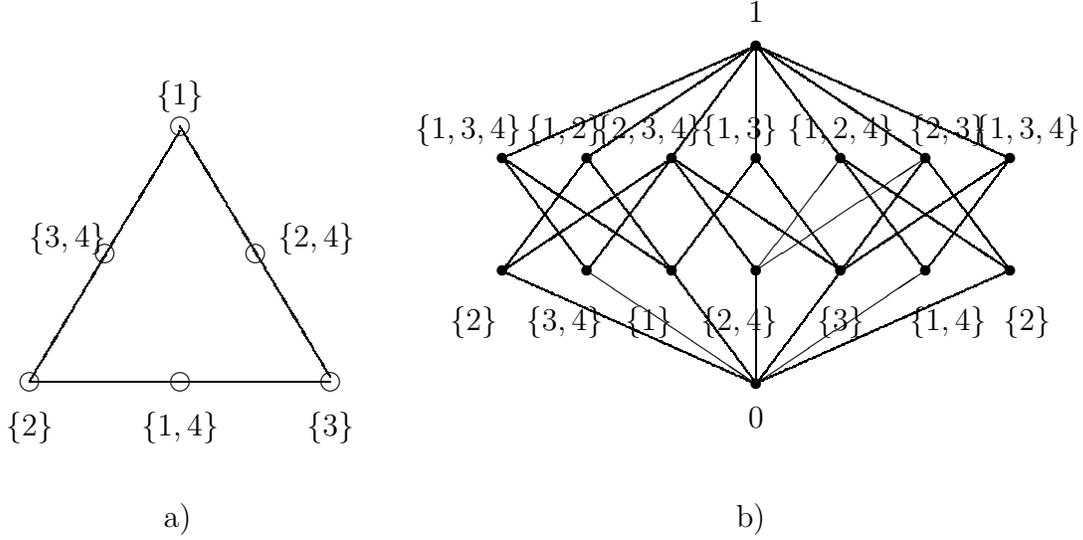

Figure
\ref{f-gh-tria} depicts the Greechie and Hasse diagrams of
this propositional structure.

The physical interpretation of Eq. (\ref{2003-cu-e-nc}) is the following:
there exist six observables
$\{1\}$, $\{2\}$, $\{3\}$, $\{1,4\}$, $\{2,4\}$, and $\{3,4\}$; i.e.,
 $\{3,4\}$ corresponds to {\it `` the system is in state $3$ or in state $4$.''}

They are grouped into three partitions of $\{1,2,3,4\}$,
such that within each group the observables are comeasurable.
For instance, in the automaton example enumerated above, the experiment
with the input of symbol 2 differentiates between
$\{1,4\}$, $\{2\}$, $\{3\}$ and all properties obtained by forming the logical
``or'' operation, such as $\{2,3\}$.
But the experiment does not reveal all conceivable
propositions, such as $\{1,2\}$.
Another experiment with the input of symbol $3$ does,  but cannot reveal other properties,
such as   $\{1,4\}$.
Because of this complementarity,
the propositions are nonclassical,
in particular they do not obey the distributive law:
since  $\{1,3\} \vee \{2,3\}= \{1,2,3,4\}$,
\begin{eqnarray*}
\{1,2,4\}\wedge (\{1,3\} \vee \{2,3\}) = \{1,2,4\} \wedge \{1,2,3,4\} &=& \{1,2,4\}
\\
=(\{1,2,4\}\wedge \{1,3\}) \vee (\{1,2,4\}\wedge \{2,3\}) &=& \{1,2\} .
\end{eqnarray*}

This humble propositional structure is thus non-classical, but
quite remarkably it also cannot
 be realized by quantum mechanics.
The complementary groups are interlocked in a triangle form, which is forbidden
by the Hilbert space based algebraic structure of quantum mechanics:
In analogy to Kochen and Specker \cite{kochen2},
we denote by the symbol ``$\perp$'' the binary relation of comeasurability.
Any sequencing of observables such as
$$\{1\}\perp \{3,4\}\perp \{2\}\perp \{1,4\}\perp \{3\}\perp \{2,4\}\perp \{1\}$$
(with
$\{1\}\not\perp  \{1,4\}\not\perp  \{2,4\}$ and so on)
cannot occur in quantum mechanics.
Hence, if this propositional structure is experienced in some physical setup,
then quantum mechanics is not an appropriate theoretical representation for it.
Computational universes would be a natural candidate.

\subsection{Coarse grained structure of digital space}

Already Zuse mentioned that, if space is tesselated, then this tesselation will
eventually show up; either by some anisotropy or by a fundamental length scale.
No indication is given exactly when this granularity should show up; and problems abound
\cite{thooft2}.
Yet, there is no guarantee that space and time will be organized
as a regular lattice;
it may rather resemble a huge pile of more or less
randomly and densely packed sand
and stabilized by whatever forces there are.

In view of the mild discreteness of quantum mechanics already mentioned
earlier (only an integer number of quanta per field node),
it might well be that we have already unravelled
the fundamental discreteness;
but not in the properties where we had expected them.
So, maybe the field nodes or phase space are more fundamental
than the frames of space and time that we use to define those fields.
In this idealistic picture,
space and time may be convenient constructions of our minds to sort
out the evolution of field modes.

\subsection{Exotic probabilities}

One approach to the formalism is that anything which is not
forbidden explicitly is realized
\footnote{
Feynman's rule of thumb states that whatever is not explicitly forbidden is mandatory.
See also the ``go-go'' principle  introduced in
\cite{svozil-set}.}.
As Gleason's theorem strongly ties quantum probabilities to
Hilbert space,
there may be non-classical and non-quantum
probabilities which can be modelled with automaton or generalized urn models.

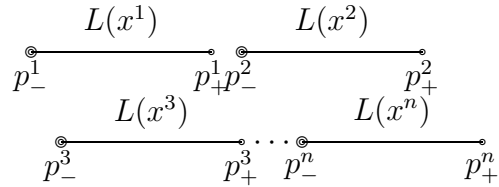
\begin{figure}[htd]
\begin{center}
%TexCad Options
%\grade{\off}
%\emlines{\off}
%\beziermacro{\off}
%\reduce{\on}
%\snapping{\off}
%\quality{0.20}
%\graddiff{0.01}
%\snapasp{1}
%\zoom{0.74}
\unitlength 0.40mm
\linethickness{0.4pt}
\begin{picture}(151.06,45.00)
\put(0.00,35.00){\circle{2.11}}
\put(0.00,35.00){\circle{3.50}}
\put(60.00,35.00){\circle{2.11}}
\put(0.00,26.37){\makebox(0,0)[cc]{$p_-^1$}}
\put(60.00,26.37){\makebox(0,0)[cc]{$p_+^1$}}
\put(0.00,35.00){\line(1,0){60.00}}
\put(70.00,35.00){\circle{2.11}}
\put(130.00,35.00){\circle{2.11}}
\put(70.00,35.00){\circle{3.50}}
\put(70.00,26.37){\makebox(0,0)[cc]{$p_-^2$}}
\put(130.00,26.37){\makebox(0,0)[cc]{$p_+^2$}}
\put(70.00,35.00){\line(1,0){60.00}}
\put(30.00,45.00){\makebox(0,0)[cc]{$L(x^1)$}}
\put(100.00,45.00){\makebox(0,0)[cc]{$L(x^2)$}}
\put(10.00,5.00){\circle{2.11}}
\put(10.00,5.00){\circle{3.50}}
\put(70.00,5.00){\circle{2.11}}
\put(10.00,-3.63){\makebox(0,0)[cc]{$p_-^3$}}
\put(70.00,-3.63){\makebox(0,0)[cc]{$p_+^3$}}
\put(10.00,5.00){\line(1,0){60.00}}
\put(90.00,5.00){\circle{2.11}}
\put(90.00,5.00){\circle{3.50}}
\put(150.00,5.00){\circle{2.11}}
\put(90.00,-3.63){\makebox(0,0)[cc]{$p_-^n$}}
\put(150.00,-3.63){\makebox(0,0)[cc]{$p_+^n$}}
\put(90.00,5.00){\line(1,0){60.00}}
\put(40.00,15.00){\makebox(0,0)[cc]{$L(x^3)$}}
\put(120.00,15.00){\makebox(0,0)[cc]{$L(x^n)$}}
\put(80.00,4.67){\makebox(0,0)[cc]{$\cdots$}}
\end{picture}
\end{center}
\caption{\label{f-gd-monm}
Example for a nongleason type probability measure
for $n$ spin one-half state propositional systems $L(x^i),
i=1,\cdots
,n$
which are not comeasurable. The superscript $i$ represents the $i$th
measurement direction.
The concentric circles indicate the atoms with probability measure 1.
}
\end{figure}

Let us consider again spin state measurements on electrons modelled by
two-dimensional Hilbert space entities.
The associated algebra of propositions consists of (the horizontal sum of)
Boolean sublattices $2^2$ which are pasted together \cite{nav:91}
at their extreme elements.
In this case, Gleason's theorem does not apply.
By taking the algebraic structure and the
set of dispersion free (two-valued) states
alone, there exists the possibility of nongleason type probability
measures.
These measures  have singular, separating distributions
and thus can be embedded into ``classical'' Boolean algebras
such as generalized urn und automaton partition logics.
One particular example is represented in Figure \ref{f-gd-monm}.
Its probability measure is $P(x_-^i)=1$ and $P(x_+^i)=1-P(x_-^i)=0$ for
$i=1,\ldots ,n$.
The associated automaton models are straightforward.
Every such dispersion free  state is obtained by associating with it a particular automaton state.
Whether or not such probability distributions exist for fundamental
processes is an open question.
For spin state measurements of the electrons, this does not seem to be the case,
but again the question of state preparation may be essential here.

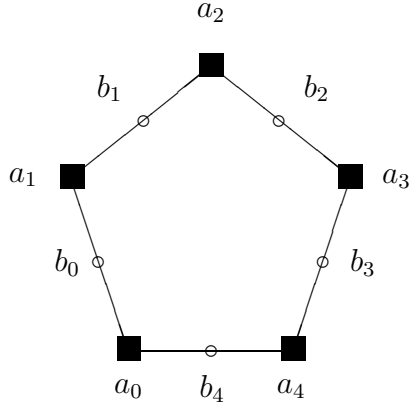
\begin{figure}
\begin{center}
%TexCad Options
%\grade{\off}
%\emlines{\off}
%\beziermacro{\off}
%\reduce{\on}
%\snapping{\off}
%\quality{0.20}
%\graddiff{0.01}
%\snapasp{1}
%\zoom{1.00}
\unitlength 0.50mm
\linethickness{0.4pt}
\begin{picture}(99.67,100.00)
\put(28.33,10.33){\circle{2.75}}
\put(72.00,10.33){\circle{2.75}}
\put(50.33,86.00){\circle{2.75}}
\put(13.33,56.33){\circle{2.75}}
\put(87.33,56.33){\circle{2.75}}
\put(13.33,56.33){\line(5,4){37.00}}
\put(50.33,85.93){\line(5,-4){37.00}}
\put(13.00,56.00){\line(1,-3){15.33}}
\put(87.33,56.00){\line(-1,-3){15.33}}
\put(72.00,10.00){\line(-1,0){44.00}}
\put(20.33,33.66){\circle{2.75}}
\put(80.00,33.66){\circle{2.75}}
\put(68.33,71.33){\circle{2.75}}
\put(32.33,71.33){\circle{2.75}}
\put(50.33,10.00){\circle{2.75}}
\put(25.33,7.67){\rule{6.33\unitlength}{6.33\unitlength}}
%{a1}}
%\put(7.00,23.66){\circle*{2.75}}%{b0}}
\put(10.33,53.33){\rule{6.33\unitlength}{6.33\unitlength}}
%{a1}}
%\put(19.00,61.33){\circle*{2.75}}%{b1}}
\put(47.33,83.00){\rule{6.33\unitlength}{6.33\unitlength}}
%{a2}}
%\put(55.00,61.33){\circle*{2.75}}%{b2}}
\put(84.33,53.33){\rule{6.33\unitlength}{6.33\unitlength}}
%{a3}}
%\put(66.67,23.66){\circle*{2.75}}%{b3}}
\put(69.00,7.67){\rule{6.33\unitlength}{6.33\unitlength}}
%{a4}}
%\put(37.00,0.00){\circle*{2.75}} %{b4}}
%\put(34.33,73.33){\rule{6.67\unitlength}{6.33\unitlength}}
\put(28.33,0.00){\makebox(0,0)[cc]{$a_0$}}
\put(50.67,0.00){\makebox(0,0)[cc]{$b_4$}}
\put(12.00,33.67){\makebox(0,0)[cc]{$b_0$}}
\put(0.33,56.33){\makebox(0,0)[cc]{$a_1$}}
\put(23.33,80.00){\makebox(0,0)[cc]{$b_1$}}
\put(50.33,100.00){\makebox(0,0)[cc]{$a_2$}}
\put(78.33,80.00){\makebox(0,0)[cc]{$b_2$}}
\put(99.67,56.00){\makebox(0,0)[cc]{$a_3$}}
\put(90.67,33.67){\makebox(0,0)[cc]{$b_3$}}
\put(71.67,0.00){\makebox(0,0)[cc]{$a_4$}}
\end{picture}
\end{center}
\caption{\label{f-lwpen}Greechie diagram of the Wright
pentagon
\protect\cite{wright:pent}. Filled squares indicate
probability~${1\over 2}$.}
\end{figure}

Another more exotic example of a suborthoposet which is embeddable into the
three-dimensional real Hilbert lattice ${\frak C}({\Bbb R}^3)$ and
can also be realized by generalized urn models and finite automata
has been presented by Wright \cite{wright:pent}. Its
Greechie diagram of the pentagonal form is drawn in Figure
\ref{f-lwpen}.
Wright showed that the probability measure
$P(a_i)={1\over 2}$, $P(b_i)=0$, for $i=0,1,2,3,4$,
as depicted in Figure \ref {f-lwpen}, is no convex combination of other
pure states; and
furthermore, that it does not correspond to any Gleason type measure
allowed as quantum probability. In this sense, it is a
``stranger-than-quantum probability.''
And although automata  and
generalized urn models ass well as quantum system
with this pentagonally interlocking algebraic structure of propositions exist,
no realizable probability measure on it is
of the form of Wright's measure defined above.
The reason for this is the impossibility to represent it
as a convex combination of other
dispersion free two-valued  states.

\subsection{``Tuning'' reality}

If the physical phenomena are the intrinsic view of a mathematical
or computational universe, then any attempt to render, manipulate and
change certain phenomena could be interpreted as ``reprogramming.''
In fact, reprogramming or ``tuning''
\footnote{
The term ``tuning'' is borrowed from the movie
{\it Dark City} by Alex Proyas, where similar motives have been casted.}
reality may be a powerful new metaphor hitherto foreign to theoretical physics.
Again, one should keep in mind that this is highly speculative.

\subsection{Against odds}
Let me again emphasise that
discrete or algorithmic physics may be  utterly non mainstream and off-topic,
as
competing with "traditional" continuum physics is hard.
For instance, note the fabulous coincidence between the theoretical
and the experimental values of the anomalous magnetic moment of the Muon
$a_{\mu ,t} = 11 659 177(7)\times 10^{-10}$ and
$a_{\mu ,e} = 11 659 204(7)(5) \times 10^{-10}$
\cite{muong-2Collaboration}.
Or take the neutron double slit experiments \cite{zeilinger:ds}
which show a wonderful agreement of theory and experiment.

Yet, despite all these difficulties,
discrete computational physics certainly represents an interesting,
speculative and challenging research area.
Many ideas from system science, interface design,
to  dualism  (e.g., the  Eccles Telegraph) enter.
The issue has metaphysical connotations.
It is for instance not totally unlikely that a demiurge would create
an "atomistic" world such as ours, in which
an immense (to us) number of discrete gaming pieces come together to form a universe and
which are constantly rearranged to
form rich and varied and seemingly complex patterns.
Or there is just one consistent Universe of Mathematics,
and this is the  physical Universe we are living in.

\section*{Acknowledgements}
Many thanks go to Tim Boykett for inviting me to a Time's Up workshop,
giving me the
feeling that somewhere out there are people still interested and listening.
Garry J. Tee from the University of Auckland has provided reference \cite{Thomas},
and Peter Mittelstaedt  referred to the beautiful quotation of Kant.
I am particularly thankful to Ross Rhodes
for his continuing encouragement; also for
reading an earlier version of the manuscript and for  many
suggestions to improve the text.

%\bibliography{svozil}
%\bibliographystyle{elsart-num}
%\bibliographystyle{apsrev}
%\bibliographystyle{unsrt}

\end{document}